\documentstyle[aps]{revtex}

\input epsf
\begin{document}
\title{Stationary dark energy : the present universe as a global attractor}
\author{Luca Amendola and Domenico Tocchini-Valentini}
\address{Osservatorio Astronomico di Roma, \\
Viale del Parco Mellini 84, 
00136 Roma, Italy}
\date{\today }
\maketitle

\begin{abstract}
We propose a cosmological model that makes a significant step toward solving
the coincidence problem of the near similarity at the present of the dark
energy and dark matter components. Our cosmology has the
following properties: a) among flat and homogeneous spaces, the present
universe is a global attractor: {\it all} the possible initial conditions lead
to the observed proportion of dark energy and dark matter; once reached, it
remains fixed forever; b) the expansion is accelerated at the present; 
c) the model is consistent with the
large-scale structure and microwave background data; d) the dark energy and
the dark matter densities  scale similarly after equivalence and are
close to within two orders of magnitude. 
\end{abstract}

\bigskip 

Since the introduction of inflationary models the notion of attractor
cosmological solutions has been regarded as a desirable property of any
successful model. Unfortunately, inflation itself has never completely
solved the problem of the initial conditions, since the subsequent
decelerated era is no longer an attractor, and any fluctuation away from
flatness will be amplified in the future, unless a new accelerated era
prevents it.

The search for cosmological attractors has been revived by the recent
findings (\cite{rie} \cite{per}) according to which the dominant component
of the universe medium is in a form of energy density possessing peculiar
characteristics: negative pressure and weak clustering. This energy, dubbed
dark energy or quintessence (\cite{fri}\cite{wet95}\cite{rat}\cite{cal}),
should fill roughly 70\% of the critical energy density and, along with
another 30\% in ordinary dark matter (and a minor component of baryons),
explains the SNIa observations, is consistent with the CMB data (see e.g. 
\cite{ame4}), and other  evidences like the cluster masses. The
fact that the energy densities of the dark energy and the dark matter are
comparable at the present time is indeed an enigma, since we have no reason
to expect that the dark energy and the dark matter components, which have
always given a very different contribute to the total density in the past
and will again give a different one in the future, are almost equal right
now. In terms of the phase-space view of the cosmological equations, the
problem is that the mixture of dark energy and dark matter we observe today
is not a global attractor; a different initial condition or, equivalently, a
different instant of observation, gives a different sharing of the total
density. The problem lies in the fact that the two energy forms scale
differently with time because they are assumed to be completely unrelated.
To explain the coincidence we propose to couple dark energy to dark matter.

The model we propose in this paper, denoted {\em stationary dark energy}, is
based on a non-linear coupling of dark energy to dark matter. The resulting
cosmological solution has the following properties: {\it a}) among flat and
homogeneous spaces, the present universe is a global attractor: {\it all}
the possible initial conditions lead to the observed percentages of dark
energy and dark matter; once reached, they remain fixed forever; {\it b})
the expansion is accelerated at the present, as requested by the SNIa
observations; {\it c}) the model is consistent with the large-scale
structure and CMB data; {\it d}) the dark energy and
the dark matter densities always scale similarly after equivalence and are
close to within two orders of magnitude. Basing on the literature known to us, no
other cosmological model satisfies all four requirements. For instance, the
quintessence model of ref. \cite{cal}\cite{zla}\cite{alb}\cite{cop2} is accelerated and
consistent with observational data but the present universe is not a global
attractor: the observed percentages of energy density will change in the
future until the cosmic medium will be dominated by quintessence alone. 
Notice
that in all these models there is a ``tracking'' solution, that can be defined
as an attractor in a subspace of the phase space; in contrast, a true attractor
as we have in our model is an attractor in the full phase space. The model
proposed in \cite{fer}, based on an exponential potential, satisfies {\it a} 
) only if is not accelerated. With the inclusion of a linear coupling
between dark energy and dark matter, as in refs. \cite{ame2}\cite{ame3}\cite
{hol}, it can satisfy {\it a}), {\it b}) and {\it d)} but still not {\it c}) 
or, alternatively, can be accelerated and consistent with observations but then
the present universe is not a global attractor. In ref. \cite{hol} a model
that can satisfy all  criteria is proposed, but it requires the
introduction of two different forms of dark matter, only one of which is
coupled to dark energy. In ref. \cite{pav} and \cite{zim} a 
dark matter with an
effective anti-friction can  satisfy {\it a}) and {\it d}) but the
effects on structure formation and microwave background have not been tested.

In our model the final state of the universe is an accelerated expansion
with a fixed ratio of dark matter and dark energy. We can remark that since
the accelerated expansion flattens a curved space, the global attractor in
our model attracts also all open universes and all those closed universes
which have not already collapsed to a singularity by the time the final
stationary state sets in (see \cite{hoo} for a discussion on attractor
solutions in curved spaces). It is to be noticed, however, that although in
our model the universe will always reach a final state that may represent
our present world, there is no guarantee that this state has already been
reached, nor that it had done so late enough to grow sufficient structure
formation. This two requirements do limit the range of acceptable initial
conditions.

In this work we introduce the concept and derive the main constraints
imposed by primordial nucleosynthesis, age of the universe, large scale
structure and SNIa results. More precise constraints from CMB observations
will be derived in another paper.
The dark energy scalar field  in our model is defined by two
functions: the coupling to dark matter and the potential. The coupling
generalizes the linear one introduced in ref. \cite{ame1} (see also \cite
{wan}\cite{hol}\cite{bil}), which is in fact the Einstein frame version of a
Brans-Dicke theory: now we will adopt a non-linear coupling, as detailed
below. To avoid the strong constraints on such a coupling we adopt the
species-dependent coupling proposed in Ref. \cite{dam}, leaving the baryons
uncoupled (see also \cite{cas}\cite{ame3}\cite{bea}). To introduce our model
we first recall the main properties of the coupled quintessence \cite{ame3}:
a scalar field with exponential potential and linear coupling. Consider
three components, a scalar field $\phi $ , baryons and CDM, described by the
energy-momentum tensors $T_{\mu \nu (\phi )},$ $T_{\mu \nu (b)}$and $T_{\mu
\nu (m)}$, respectively. General covariance requires the conservation of
their sum, so that it is possible to consider a coupling such that 
\begin{eqnarray}
T_{\nu (\phi );\mu }^{\mu } &=&\left( C_{m}T_{(m)}+C_{b}T_{(b)}\right) \phi
_{;\nu },  \nonumber \\
T_{\nu (m);\mu }^{\mu } &=&-C_{m}T_{(m)}\phi _{;\nu },\quad T_{\nu (b);\mu
}^{\mu }=-C_{b}T_{(b)}\phi _{;\nu }.  \label{coup1}
\end{eqnarray}
A similar coupling is obtained by conformally transforming a non-minimally
coupled gravity theory. The radiation field (subscript $\gamma $) remains
uncoupled, since $T_{(\gamma )}=0.$ We derive the background equations in
the flat FRW metric, assuming the exponential potential 
\begin{equation}
U=Ae^{s\phi }
\end{equation}
as proposed e.g. in \cite{rat}\cite{wet95}\cite{fer}. As anticipated, we
will couple the dark energy scalar field to the dark matter only, putting $ 
C_{b}=0$. We call this choice dark-dark coupling. Introducing the  variables 
\cite{cop} $x=\frac{\kappa }{H}\frac{\dot{\phi}}{\sqrt{6}},\quad y=\frac{ 
\kappa }{H}\sqrt{U/3},\quad z=\frac{\kappa }{H}\sqrt{\rho
_{\gamma }/3}$, (where $H=\dot{a}/a$ , $G=c=1$ and $\kappa ^{2}=8\pi $),
and adopting the $e$-folding time $\alpha =\log a$, we can write the field
and radiation conservation equation as a system in the variables $x,y,z$
that depends on the parameters $\mu ,\beta $ 
\begin{eqnarray}
x^{\prime } &=&\left( z^{\prime }/z-1\right) x-\mu y^{2}+\beta
(1-x^{2}-y^{2}-z^{2}),  \nonumber \\
y^{\prime } &=&\mu xy+y\left( 2+z^{\prime }/z\right) ,  \nonumber \\
z^{\prime } &=&-z\left( 1-3x^{2}+3y^{2}-z^{2}\right)/2 ,
\label{sys1}
\end{eqnarray}
where the prime denotes derivation with respect to $\alpha $, and where $ 
\beta =C_{m}\sqrt{3/(2\kappa ^{2})},\quad \mu =s\sqrt{3/(2\kappa
^{2})}.$ The baryons are here neglected, since they act on the dynamical
system only as a minor perturbation. The dimensionless constant $\beta $
sets the ratio of the strength of the dark-dark interaction with respect to
the gravitational interaction; $\beta $ is clearly not constrained by local
experiments or by $\dot{G}/G$ measurements \cite{dam}. Here we restrict the
attention to $\beta >0$, $\mu >0$, without loss of
generality (see \cite{ame3}). 

The system (\ref{sys1}) has several different global attractors, depending
on the values of the parameters $\beta $ and $\mu $, but only two can be
accelerated. One, to be denoted attractor $a$, exists for $\mu <3$ , and is
accelerated for $\mu <\sqrt{3}$. On this attractor, the energy density is
entirely in the dark energy component, and as such it cannot represent our
universe. Of course, our universe could be described by this solution if the
attractor has not been reached yet (see e.g. \cite{ame4}).
The other, the attractor $b$, exists for $\mu +\beta >3/2$ and is
accelerated for $\mu <2\beta $ . On this attractor, contrary to the case $a$ 
, the energy density is shared by the dark energy and the dark matter in the
following proportions 
\begin{equation}
\Omega _{\phi }=\frac{4\beta ^{2}+4\beta \mu +18}{4(\beta +\mu )^{2}},\quad
\Omega _{m}=1-\Omega _{\phi }  \label{attrb}
\end{equation}
The universe expands as a power law $a\sim t^{p}$ with exponent 
$
p=2( 1+\beta /\mu )/3\,. 
$
Choosing the parameters we can ensure that the final state is, for instance, 
$\Omega _{\phi }=0.7$ and $\Omega _{m}=0.3$. Once reached, these values will
remain fixed forever. The problem with the model above is that when the
radiation epoch ends, the system reaches rapidly the attractor $b$ and a
matter dominated epoch never sets in: the inhomogeneities never grow and the
model fails completely to explain the large scale structure. What is lacking
is an intermediate phase of matter domination and structure formation. Such
a phase would be present if $\Omega _{\phi }$ were very small, say less than
0.1, so that $\Omega _{m}$ dominates. However, as can be seen from Eq. (\ref
{attrb}), to get a small $\Omega _{\phi }$ we need a large ratio $\mu /\beta 
$, but then this attractor would not be accelerated since $p<1$. In other
words, if we want acceleration, we need a large coupling, $\beta \gg \mu $;
if we want structure formation, we need on the contrary a small coupling, $ 
\beta \ll \mu $. To have both, we need two couplings.

To realize this we use a non-linear coupling that switches between a small
(or zero) $\beta _{1}$ and a large $\beta _{2}$ when $\phi $ rolls down the
potential. For the numerical calculations of this paper we assume
\begin{equation}
\beta (\phi )=\frac{1}{2}\left[ (\beta _{2}-\beta _{1})\tanh \left( \frac{ 
\phi _{1}-\phi }{\Delta }\right) +\left( \beta _{2}+\beta _{1}\right) \right]
.  \label{tanh}
\end{equation}
The precise form of the coupling function $\beta (\phi )$ is not really
important; any step-like function that switches on the coupling after
structure formation will give qualitatively similar behavior.  We choose the constants $\beta
_{1}$ and $\beta _{2}$ and the slope $\mu $ so that 
\begin{equation}
\beta _{1}\ll \mu \ll \beta _{2}.  \label{ineq}
\end{equation}
The dynamics of the model can then be summarized as follows. When
the universe leaves the radiation era, if $\phi >\phi _{1}$, the coupling is
effectively $\beta _{1}$, and the system falls on the attractor (\ref{attrb}
) with $\beta =\beta _{1}$. Taking a small $\beta _{1}$ and a large $\mu $,
the contribution of $\Omega _{\phi }=9/2\mu ^{2}$ is small and the matter
dominated era allows the inhomogeneities to grow. Moreover, the condition $
\mu \gg 1$ will ensure that the primordial nucleosynthesis is not
excessively altered. When $\phi $ rolls below $\phi _{1}$, the coupling
becomes stronger and the final global accelerated attractor (\ref{attrb})
with $\beta =\beta _{2}$ is reached. The condition $\beta _{2}/\mu \gg 1$
(actually $\beta >2\mu $ is sufficient) ensures that the second and last
power law is accelerated. The universe will inflate forever with a constant
ratio of dark energy to dark matter. The value $\phi _{1}$ sets the instant
at which the coupling changes strength and the universe crosses from the
dark matter epoch to the dark energy epoch, while $\Delta $ modulates the
rapidity of the transition. For the model to explain the large scale
structure, it is crucial that this transition occurs late enough for the
inhomogeneities to grow.

With the coupling (\ref{tanh}) we obtain the same equations (\ref{sys1})
where now the constant $\beta $ becomes a function $\beta (y,H)$ and where
an extra equation for $H$ is needed,
$
H^{\prime }=-\frac{3H}{2}( 1+x^{2}-y^{2}+z^{2}/3) .
$
In Fig. 1 we present a numerical integration of the full set of equations. As
expected, there exist three distinct phases of constant energy density
ratios among the various components. First, the model passes through a
radiation dominated epoch with a vanishing contribution of matter and a
small or vanishing contribution of the scalar field. After equivalence, it
falls upon the saddle (\ref{attrb}) where, for small $\beta _{1}$, matter
dominates but there is also a finite contribution from the dark energy ($ 
y=-x=3/2\mu $ in the limit of small $\beta _{1}$): this stage is denoted
plateau I. Finally, it enters the present epoch of dark energy domination,
with a 30\% contribution from the dark matter: this is the plateau II. 
 In the same Fig. 1 we plot
also the effective parameter of state $w_{eff}=1+\frac{p_{tot}}{\rho _{tot}}$ 
: values $w_{eff}<2/3$ imply acceleration. 

An obvious objection to our model is that we are trading the coincidence
between dark energy and dark matter for a coincidence with the instant when
the strong coupling is switched on. However, it is to be remarked that in
our case dark energy and dark matter have been similar to within one or two
orders of magnitude ever after equivalence ($\Omega _{\phi }$ goes from $ 
9/2\mu ^{2}$ in plateau I to 0.7 in plateau II), so that the present
coincidence is no longer particularly striking: in other words, after
equivalence, the ratio of dark matter to dark energy is never really far
from unity, while in all the other models it is so only at one particular
instant, and extremely large or small at any other time. In addition, the
fact that dark energy and dark matter are allowed to reach a constant
proportionality only after equivalence explains also the ``triple
coincidence'' noticed by Arkani-Hamed et al. \cite{ark} among radiation and
the other components: in our model, it is the end of radiation dominance
that triggers dark energy and dark matter to equalize. In Fig. 2 we contrast
the behavior of the ratio $\rho _{\phi }/\rho _{m}$ in our model (for
different initial conditions) with that in a inverse power-law as in  \cite
{cal}. As can be seen, while in our case the ratio remains relatively close
to unity at all times after equivalence, in the inverse power-law case this
coincidence occurs only today.

Let us consider now the main constraints on the model. First of all, we fix $ 
\Omega _{\phi }=0.7\pm 0.1$, and $w_{eff}|_{0}\in (0,0.6)$ as required by
SNIa observations along with the condition of flatness. Then, we impose that
the universe age be sufficiently large. Neglecting the radiation epoch and
assuming instantaneous transition from plateau I to plateau II at a redshift 
$z_{c}$ we obtain 
\begin{equation}
T=\frac{2}{3H_{0}}\left[ \frac{1-(1+z_{c})^{-\frac{3}{2}w_{2}}}{w_{2}}+\frac{ 
(1+z_{c})^{-\frac{3}{2}w_{1}}}{w_{1}}\right] ,
\end{equation}
where $w_{1}=\mu /(\mu +\beta _{1})$ and $w_{2}=\mu /(\mu +\beta _{2})$.
Because of the inequalities (\ref{ineq}) we may approximate, if $z_{c}\gg 1$
(but also $z_{c}\ll 1000$ in order to ensure structure formation; $ 
z_{c}\simeq 5$ would be acceptable) $T=\frac{2}{3H_{0}}(\frac{3}{2}\log
z_{c}+z_{c}{}^{-\frac{3}{2}}),$ which is always larger than the matter
dominated age $2/(3H_{0})$, so that we pass easily the age test. 

We come now to the condition of sufficient structure formation. The growth
of perturbations in the plateau I has been considered in Ref. \cite{ame2}.
The dark matter inhomogeneity $\delta $ during the plateau I grows as $a^{m}$
with $m=\frac{1-p_{1}}{2p_{1}}[-1+\left[ 1+F(\Omega _{\phi 1},p_{1})\right]
^{1/2}]$ , $p_{1}$ and $\Omega _{\phi 1}$ being the scale factor exponent
and the field energy density parameter during the plateau I and where 
\[
F(\Omega _{\phi },p)=\frac{6p(1-\Omega _{\phi })\left[ -8+26p+3(\Omega
_{\phi }-7)p^{2}\right] }{(p-1)^{2}\left[ 2+3p(\Omega _{\phi }-1)\right] }. 
\]
This reduces to the usual linear growth $m=1$ for $p_{1}=2/3$ and $\Omega
_{\phi 1}=0$, that is in the standard case without scalar field. In all the
other cases, $m<1$. Therefore, in order for the perturbation to grow not
much less than in the usual matter dominated era, we need to be close to the
standard case. For instance, if $m=0.9$, then the fluctuations grow from $ 
z\simeq 1000$ down to $z\simeq 1$ by half the standard case. Considering the
present uncertainty on the amplitude of fluctuations at the present, we may
take this as the lower limit for the fluctuation growth. Requiring also $ 
\beta >0,m>0.9$ implies $\mu >28.1$.
Finally, the nucleosynthesis constraint reduces to the request that during
the radiation era the contribution of the dark energy is sufficiently low,
e.g. less than 15\%. During the radiation era the system passes through one
of two saddles, depending on the initial conditions. In one, labelled $b_{R}$
in Ref. \cite{ame3}, there is a constant contribution from the scalar field,
corresponding to $\Omega _{\phi }=6/\mu ^{2}$: 
nucleosynthesis requires 
 $\Omega _{\phi }\leq 0.15$ ( see
e.g. \cite{fer}) or  $\mu >7$. In the second saddle the
contribution from the field vanishes, so there are no additional constraints
from it. In Fig. 3 we summarize the constraints derived so far. Every
coupling function, or potential, that moves the effective parameters $\mu
,\beta $ from region I (the gray region on the left) to region II (the gray
region on the right) after structure formation produces an acceptable model.

In conclusion, we have shown that it is possible to construct a relatively
simple model in which the present universe has {\it already} reached the
global attractor. This offers two advantages over the previous dark energy
models with tracking solutions. The first is that the presently observed
ratio of dark matter to dark energy density has been close to within two
orders of magnitude ever after equivalence, thereby reducing the impact of
the cosmic coincidence problem. The second is that {\it all} the initial conditions
will lead sooner or later to this state, while in all the other models only
a finite fraction of the phase space lead to our universe (see e.g. the
discussion in \cite{zla}). The near coincidence of
dark energy and dark matter energy densities does not longer depend on the
initial conditions but only on the coupling constants, and will be the same
at any future epoch. We believe this is a significant step toward the
solution of the cosmic coincidence problem. Of course, current observations
do delimit the range of acceptable initial conditions: in fact, many initial
conditions will give trajectories that fall onto the final attractor either
too soon, so that not enough structure forms, or too late, so that we are
still short of the attractor. In other words, although the phase-space
trajectory that the universe follows from some point onward is unique and
independent of the initial conditions, the current position on the
trajectory do depend on them, as in all cosmological models.

{\it Aknowledgments}. L.A. thanks David Wands for useful discussions and for
the hospitality at the University of Portsmouth where part of this work has
been carried out.

\bigskip

\bigskip

\newpage

\begin{figure}[tbp]\epsfysize 8in
\epsfbox{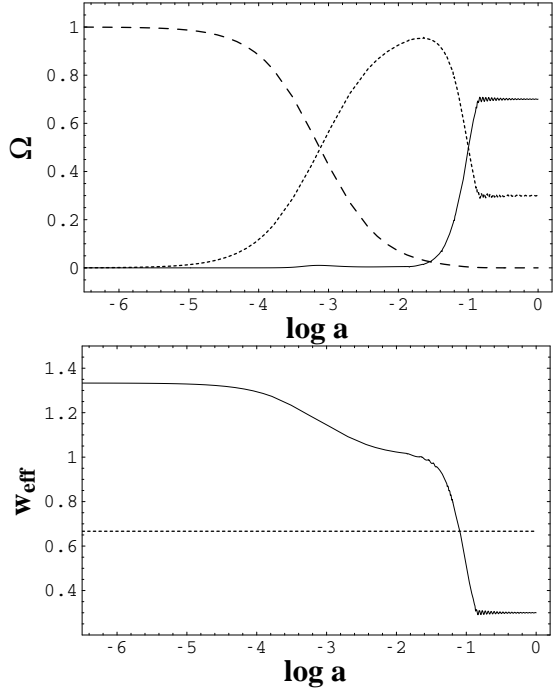}
\caption{ Top panel. Trends of $\Omega _{\protect\gamma }$ (dashed line), $ 
\Omega _{m}$ (dotted), and $\Omega _{\protect\phi }$ (continuous) versus $ 
\log a$. The three regimes mentioned in the text are evident: first,
radiation dominates, then matter dominates (plateau I), and then finally the
system falls on the final accelerated attractor (plateau II) with 30\% of
dark matter and 70\% of dark energy. The constants have been chosen here as $ 
\protect\mu =30$ , $\protect\beta _{1}=0$ and $\protect\beta _{2}=70$.
Bottom panel. The effective parameter of state $w_{eff}$ during the three
regimes: first equals $4/3$, then goes down to $1,$ and finally becomes
accelerated, $w_{eff}=0.3$.}
\end{figure}
\newpage

\begin{figure}[tbp]\epsfysize 8in
\epsfbox{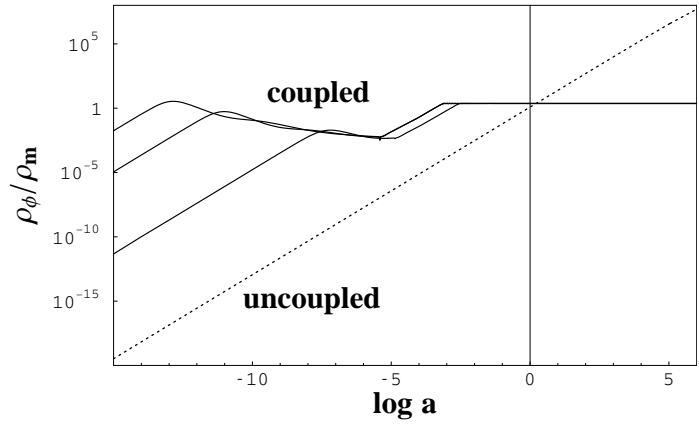}
\caption{ The plot shows the behavior of $\rho _{\protect\phi }/\rho_m$ in our model (continuous lines) for different initial conditions, and in an inverse power-law model without coupling (dotted line). The vertical line is the present time. In the coupled model the ratio is close to unity ever after equivalence.}
\end{figure}
\newpage

\begin{figure}[tbp]\epsfysize 8in
\epsfbox{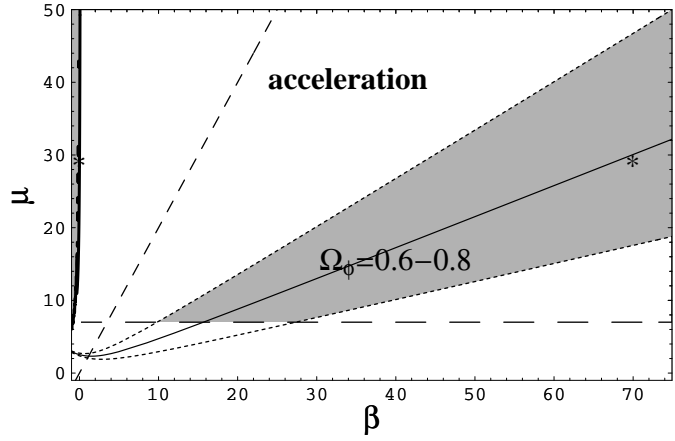}
\caption{ Parameter space of the model. To the right of the short-dashed
line the expansion is accelerated; above the long-dashed line the
nucleosynthesis constraint is passed. The parameters within the gray region
on the left produce enough structure formation. Those inside the gray region
on the right yield an accelerated expansion with $\Omega _{\protect\phi }$
between 0.6 and 0.8 (the continuous line is $\Omega _{\protect\phi }=0.7$).
Any coupling function that switches from the first region to the second
after structure formation gives an acceptable \ model. The two asterisks
mark the effective parameters we employed in the numerical calculations.}
\end{figure}
\newpage

\bigskip

\end{document}